\documentclass[12pt,showpacs,preprintnumbers,floatfix]{article}
\usepackage{graphicx} 
\usepackage{dcolumn} 
\usepackage{bm} 
\usepackage{amsfonts}
\usepackage{amsthm}
\usepackage{amsmath}
\usepackage{amssymb}
\usepackage{epsfig}
\usepackage{color}
\usepackage{textcomp}
\usepackage{hyperref}
\usepackage{titlesec}

\def\be{\begin{equation}}
\def\ee{\end{equation}}
\def\ba{\begin{eqnarray}}
\def\ea{\end{eqnarray}}

\def\bdm{\begin{displaymath}}
\def\edm{\end{displaymath}}

\def\bq{\begin{quote}}
\def\eq{\end{quote}}

 at 10truept

\newcommand{\bea}{\begin{eqnarray}}
\newcommand{\eea}{\end{eqnarray}}

\newcommand{\bi}{\begin{itemize}}
\newcommand{\ei}{\end{itemize}}

\newcommand{\beq}{\begin{equation}}
\newcommand{\eeq}{\end{equation}}
\newcommand{\beqa}{\begin{eqnarray}}
\newcommand{\eeqa}{\end{eqnarray}}

\def\ltap{\ \raise.3ex\hbox{$<$\kern-.75em\lower1ex\hbox{$\sim$}}\ }
\def\gtap{\ \raise.3ex\hbox{$>$\kern-.75em\lower1ex\hbox{$\sim$}}\ }
\def\gl{\ \raise.5ex\hbox{$>$}\kern-.8em\lower.5ex\hbox{$<$}\ }
\def\roughly#1{\raise.3ex\hbox{$#1$\kern-.75em\lower1ex\hbox{$\sim$}}}

\begin{document}

\thispagestyle{empty}
\begin{flushright}
April 2016
\end{flushright}
\vspace*{1.7cm}
\begin{center}
{\Large \bf The Fine Structure Constant and Habitable Planets}\\

\vspace*{1.2cm} {\large McCullen Sandora\footnote{\tt
sandora@cp3.dias.sdu.dk}}\\
\vspace{.5cm} {\em CP$^3$-Origins, Center for Cosmology and Particle Physics Phenomenology \\ University of Southern Denmark, Campusvej 55, 5230 Odense M, Denmark}\\

\vspace{2cm} ABSTRACT
\end{center}
We use the existence of habitable planets to impose anthropic requirements on the fine structure constant, $\alpha$.  To this effect, we present two considerations that restrict its value to be very near the one observed.  The first, that the end product of stellar fusion is iron and not one of its neighboring elements, restricts $\alpha^{-1}$ to be $145\pm 50$.  The second, that radiogenic heat in the Earth's interior remains adequately productive for billions of years, restricts it to be $145\pm9$.  A connection with the grand unified theory window is discussed, effectively providing a route to probe ultra-high energy physics with upcoming advances in planetary science.

\vfill \setcounter{page}{0} \setcounter{footnote}{0}
\newpage

\section{Introduction}

It has been a long-standing goal of theoretical physics to explain as much as possible about the universe around us from first principles.  To date, this enterprise has culminated in the standard model plus gravity, a framework that is extraordinarily successful at describing the underpinnings of practically every natural phenomenon we encounter.  Nevertheless, this theory is viewed as incomplete:  it has many free parameters, such as the masses of particles and strength of forces, etc., that cannot be explained.  It could be that the final extension of the standard model will uniquely fix all parameters, providing an ultimate reason for why nature had to turn out the way we see it.  However, this may also not be the case, and at least some of the currently observed parameters may in principle be allowed to take on other values.  This is the multiverse hypothesis, and it quickly runs into complications due to our ignorance of which parameters should be allowed to vary, the ambiguity of the relative probabilities of any of these values, necessary selection effects that the parameters we observe must take values suitable for complex life, and our woefully inadequate understanding of what forms complex life may even take, to list a few.

    One attempt to uncover evidence for the existence of other possible universes with different values for a given parameter is to utilize ``the principle of living dangerously'' \cite{just so}:  that is, if a slight change in the value of this parameter would result in an inhospitable universe, we may infer that this parameter is allowed to scan in some sort of ensemble of universes. This reasoning demotes apparent coincidences to the selection effect that we happen to live in a realization where life is possible.  Coincidences explained by this principle cannot be taken as direct evidence of a multiverse, but they serve as a useful proxy in the absence of immediate prospects for experimental verification.  As we will discuss below, its application can even constrain the types of completions of the standard model we should be willing to consider, by demanding that certain features be satisfied.  

In this paper, we apply this reasoning to the fine structure constant, $\alpha=e^2/\hbar c=1/137.036=.0073$.  Though it is one of the most well known constants, up to this point anthropic bounds on this constant have been incomplete, a consequence of the fact that traditional anthropic considerations about stars and nuclei have only been able to constrain it in combination with other fundamental variables.  The aim of this paper is to use the existence of habitable planets to place further boundaries in the space of allowed physical parameters, that lift this degeneracy and pinpoint the observed value to a small window being compatible with life (taken to mean life as we know it throughout).  We do this by considering several different processes in the universe, and deducing their effects on properties of planets.

There is some precedent for combining the fields of planetary science, fundamental physics, and anthropics.  Half a century ago, geological observations \cite{earf} were used to place constraints on Dirac's large number hypothesis \cite{numbers}, which posited a decrease in Newton's constant to explain the several appearances of $10^{40}$ in dimensionless ratios of constants. (Preceded by an even earlier period, when this was used to explain continental drift and the faint young sun paradox, before plate tectonics and the greenhouse effect became accepted explanations \cite{dicke1}.)  Ultimately, the anthropic requirement that habitable planets must have time to develop resolved the large number puzzle \cite{dicke2}.  The idea that Earth's radiogenic heat may be anthropically selected was speculated upon in \cite{just so}.  In \cite{coin}, the requirement of habitable planets was used to alleviate the coincidence problem, that the dark energy abundance is so close in magnitude to the matter density at the present time only.  Recently, several criteria for planets to even vaguely resemble the ones we are familiar with were laid out in \cite{standp}.

However, historically the interface between these fields has been extremely sporadic.  This is, no doubt, because many aspects of planetary science remained inaccessible for many years.  This was due to both the inherent complexity of the processes involved, as well as the paucity of data needed to test our theories.  This past decade has seen remarkable advances in knowledge not just of our own solar system, but of others as well, making this the right time to revisit the role of planetary habitability in anthropic selection effects.

Another reason for the lack of attention planets received in anthropic reasoning is the sheer diversity of planets observed in our universe.  True, the conditions for habitability are very restrictive, but given the large spread in characteristics, it may seem that at least some planets would by chance be suitable for life.  We point out that this philosophy is misguided, given the multiple different conditions needed for habitability.  To give just two, a planet must have the right mass in order to maintain a suitable atmosphere (water vapor but not helium), while also having the right mass in order to support plate tectonics (not molten and not a ``stagnant lid").  The conditions for both of these are very narrow, and there is no {\it a priori} reason that they should overlap.  Yet, in our universe, they do.  Explorations of relationships like this (and the subtleties involved) are the subject of this paper.

The first process we consider is the end product of stellar fusion, which results in the heaviest element for which this process is exothermic.  In our universe, the chief byproduct of this abrupt halt is iron 56, which in turn is ten to one hundred times more abundant than the neighboring metallic elements on the periodic table.  This is a consequence of it having the highest binding energy, subject to some availability constraints that we discuss below, which is the outcome of a balance between Coulomb forces and Fermi degeneracies within the nucleus.  We show that if the fine structure constant goes beyond the region $1/205<\alpha<1/95$, then iron would not be the most abundant of this class, but would instead be replaced by either nickel or chromium.  This is done in Section \ref{iron}.

What exactly is the importance of iron for life?  There are several possible answers.  The first is the prime role it plays as the Earth's core, which generates the magnetic field shielding us from harmful cosmic rays.  A second observation is that iron plays a pivotal role in every organism, both in oxygen transport and in DNA synthesis, and iron availability can even be the limiting factor in population sizes for microorganisms.  Third, there are multiple proposals that iron played a role in the early development of life.  Evidence for each of these are elaborated in more detail in section \ref{life}, and several ways of distinguishing between these hypothesis will be laid out in the conclusions.

Even more stringent bounds are placed by considering the dependence of the Earth's radiogenic heat in section \ref{radio}.  This is important for life because this provides nearly half of the Earth's internal heat, powering plate tectonics, the process that continuously delivers the raw nutrients necessary for life to the Earth's surface.  Because this relies on decays of heavy radioactive nuclei, the heat generated depends exponentially on the fine structure constant, and demanding that plate tectonics be operational leads to the window $1/153<\alpha<1/136$.  Notice that the observed value is very close to the upper bound coming from this, which is what we would expect if the principle of living dangerously were operational if there is a strong preference for large $\alpha$.  Section \ref{radiolife} expands on this result and its relevance for life.

\begin{figure*}[h]
\centering
\includegraphics[width=12cm]{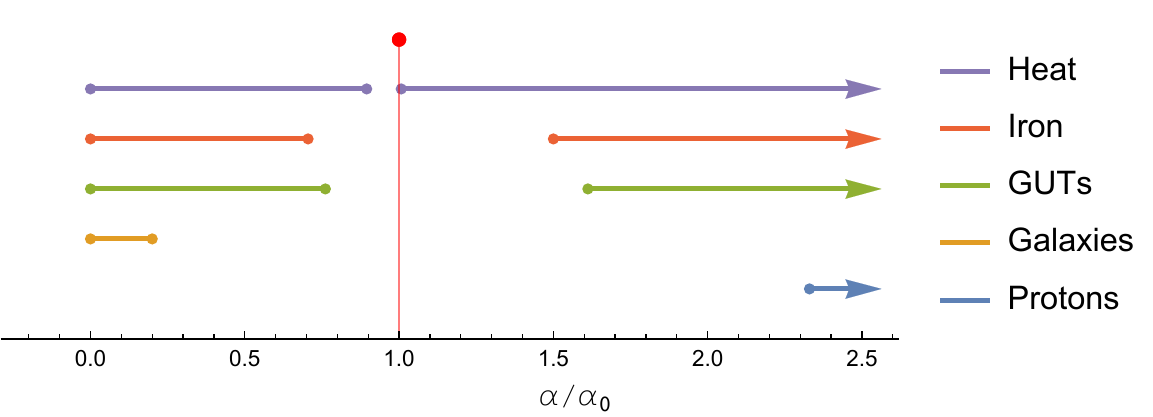}
\caption{The allowed windows of $\alpha/\alpha_0$ for the various considerations used in the text, where $\alpha_0$ is the observed value.  The bottom two are previous bounds based off the stability of protons and the cooling of galaxies, respectively.  The third is the region allowed if grand unified theories are true.  The fourth and fifth are derived in this work, and are based off the abundance of iron and the level of radioactivity necessary for Earth to be habitable.}
\label{window}
\end{figure*}

Lastly, in section \ref{nitty}, we comment on the other restriction of the fine structure constant that has appeared in the literature, that coming from grand unified theories.  GUTs typically restrict the range to be within $1/180<\alpha<1/85$, though the precise numbers depend on the specific unified theory.  This presents a puzzle:  if there are two potential reasons for the observed value of $\alpha$, one being that we rely on it for existence, and the other being that it is a necessary consequence of theories of a given type, it would be extraordinarily fortuitous if both reasons were correct.  We present several lines of reasoning along this path in section \ref{nitty}, and conclude that this leads us to expect the GUT theory to have a free parameter capable of altering the low energy coupling constants.  For orientation, we display all the bounds on $\alpha$ that we discuss in Fig. \ref{window}.

\section{Forces and Anthropics}\label{forces}

Physicists have been unable to uniquely determine the fine structure constant using anthropic arguments thus far, which is perhaps not surprising, given that there are a greater number of fundamental variables than anthropic considerations.  This work does not uniquely pin down $\alpha$ either, if other constants are allowed to vary, but it lifts the degeneracies with previous variables, and provides additional constraints in the parameter space of allowable theories.  Early work was only able to draw the conclusion $\alpha<1/6$, based on the fact that elements beyond $Z>1/\alpha$ are in the nonperturbative regime, and that a larger value of $\alpha$ would probably preclude the existence of carbon \cite{bt}.  Insights from ab initio simulations \cite{ksas} indicate that unless $\alpha<1/20$, the dipole nature of water molecules would be significantly diminished, possibly affecting its properties as a universal solvent.  In \cite{hn}, it was found that, holding all other constants fixed, requiring that the proton be stable demands $\alpha<1/59$.  If quark masses and the strong coupling scale of quantum chromodynamics (QCD) are also allowed to vary, this bound becomes a function of these variables as well.  It is even possible to place a lower bound on the strength of electromagnetism, based off the consideration that galaxies must be able to cool efficiently or else they will not collapse to form stars.  In \cite{schellekens}, the bound $1/685<\alpha$ was found, based off the formalism of \cite{tr}.  There is one analysis \cite{page} that claimed to pin down the fine structure constant to within $3\%$, but this result is based off theoretical assumptions that may or may not be valid.

So far, the tightest relationships have been arrived at by considering the processes occurring in stars.  First, it has been noticed \cite{pl} that the spectral temperature of stars in our universe happens to exactly coincide with typical molecular binding energies.  This coincidence is reliant on the apparent conspiracy between parameters
\beq
\alpha^{12}\left(\frac{m_e}{m_p}\right)^4\approx \frac{m_p^2}{M_{Pl}^2}
\eeq
Where $m_e$, $m_p$, and $M_{Pl}$ are the electron, proton and Planck masses.  If this condition did not just so happen to be satisfied, stars would either emit sterilizing ultraviolet light or faint infrared light, incapable of eliciting photosynthetic life.

Further constraints can be found by considering the Hoyle resonance in carbon.  This is a miraculous coincidence among an excited state of carbon and the energy of the beryllium-8$+$helium-4 system.  This creates a resonance in this production channel that significantly enhances the efficiency of carbon production, leading to much more carbon in our universe than would be there without.  Demanding this be present leads to the condition that $\alpha$ cannot deviate by more than $2-3\%$ \cite{ocs}.  However, if one considers simultaneous variation of $\alpha$ and the average light quark mass, there is a one-parameter line of degeneracy that would keep the abundance fixed, dependent on modelling assumptions on how the width and energy of the resonance depend on these parameters\cite{ekllm}.  These dependencies can be approximately captured by considering various models of nuclei, of which a few were chosen, that all lead to similar results.

Tying the fine structure constant to the light quark masses may seem to determine it completely, especially since the latter dictates the strength of the strong force between nucleons through pion exchange. Indeed, a minute variation of the strength of the strong force would alter the stability of the lightest nuclei, which would greatly affect properties of our universe \cite{bt,rees}:  Though the exact numbers depend on modelling assumptions for nuclear dependence on fundamental parameters, if the strong force were stronger by several percent, the diproton would be stable, providing a route by which to circumvent the deuterium bottleneck in stars, and significantly shortening their lifetime (though see \cite{dipins} for a dissenting analysis on this matter).  Likewise, if the strength were several percent weaker, deuterium would be unbound, and stars would be unable to burn their fuel at all.  Changing the fine structure constant has no effect on this, since making it stronger will not unbind the deuteron, and, even taking it to 0, the diproton would remain unbound \cite{bradford}.  This does still not determine $\alpha$, however, as these properties were shown in \cite{df} to only depend on the composite parameter $c=m_\pi/\Lambda_{QCD}$ which dictates the range of pion exchange ($\Lambda_{QCD}$ is the energy scale at which quantum chromodynamics becomes strongly coupled).  Thus, any change in the light quark masses may be compensated by a change in $\Lambda_{QCD}$ to retain dinulceon stability properties.  Changing the QCD scale will in turn affect the proton and neutron masses, but this can also be compensated by changing the strange quark Yukawa coupling \cite{dinter}, leaving enough free parameters to satisfy all constraints given here.  The general lesson is that there are many more parameters than anthropic constraints, so they are not yet completely determined by these considerations alone.

\section{The Iron Peak}\label{iron}

We now discuss how a change to the fine structure constant would affect the abundances of the transition metals.  The intermediate mass elements, including the transition metals we will be interested in, are all produced by stellar fusion, and only afterwards are distributed throughout the universe by supernovae.  Stellar evolution converts initial hydrogen and helium into heavier elements as follows (e.g. \cite{iliad}): After an initial protracted phase of converting almost all primordial hydrogen into helium nuclei, the primary fusion chain combines three heliums together to form carbon, using the Hoyle resonance to utilize the unstable beryllium before it decays.  Once there is an abundance of carbon, any leftover hydrogen can be converted to helium through the CNO cycle, providing an additional route for hydrogen burning.  This results in nuclei on the alpha ladder, that is, with the same number of protons and neutrons, and both even.  This is the preferred ratio for small elements where the Fermi energy is the dominant contribution, but which tend to be less stable once the transition metals are reached.  After all hydrogen is exhausted, fusion can only take place by climbing this ladder, either one rung at a time by helium accretion, or, in more massive stars, by directly fusing heavy elements like carbon, oxygen and silicon together.  Either way, in a very short timeframe compared to the lifetime of stars, this process occurs until the binding energy ceases to be positive, which, as in accordance with the nuclear shell model, occurs at nickel-56, which has a ``doubly magic'' number of nucleons.  This is not unstable enough to eject any nucleons through proton or $\alpha$ decay, but it migrates towards the stable elements through $\beta$ decay, first through cobalt-56 in a matter of about a week, then to iron-56 several months after that.  This results in the majority of the iron that we observe around us today.  

Thus, to understand how this process might change in universes with different coupling constants, we need to determine the conditions for stability of the various nuclei of atomic mass 56.  To do this, we use the semi-empirical (Bethe-Weizs{\"a}cker) mass formula \cite{krane}, which, while being phenomenological, has the benefit that the dependence on the underlying constants is easily traceable, and is actually remarkably successful at describing properties of nuclei.  The model posits that the binding energy of a given nucleus with total number of nucleons $A$ and charge $Z$ is
\beq
E_b(A,Z)=a_vA-a_sA^{2/3}-a_c\frac{Z(Z-1)}{A^{1/3}}-a_{sym}\frac{(A-2Z)^2}{A}+\delta
\eeq
The first terms are the volume and surface energies, which result from an interplay between nearest neighbor strong interactions and Fermi repulsion, but which are ultimately only dependent on $A$, so will be unimportant for our purposes.  The third term is due to Coulomb repulsion, the fourth gives preference to nuclei with equal numbers of protons and neutrons due to Fermi repulsion, and the fifth gives preference to even numbers of each nucleon type, encapsulating spin coupling.

Next, we remark on how the parameters of this model scale with the constants.  The Coulomb energy is directly proportional to $\alpha$, and is evaluated at a radius set by the Compton wavelength of the proton, so will be proportional to $m_p$ as well.  The symmetry term is driven by the Fermi energy, and so is only proportional to $m_p$.  Since this is a semi-empirical model, the overall coefficients must be set by experiment, though they are qualitatively fairly well understood.  We adapt the numbers from \cite{krane} to include dependence on $\alpha$ and $m_p$:
\beq
a_c=.72 \frac{\alpha}{\alpha_0}\frac{m_p}{m_{p0}}\text{MeV}, \quad a_{sym}=23\frac{m_p}{m_{p0}}\text{MeV},\quad \delta=34\frac{(-1)^Z}{A^{1/2}} \frac{m_p}{m_{p0}} \chi_{evens}(A) \text{MeV}
\eeq
The function $\chi$ is the indicator function, which vanishes unless $A$ is even, in which case it equals 1.  Changing the fine structure constant will never make the odd-odd elements more stable than the even-evens, but to $\beta$ decay the nuclei must pass through the odd-odd elements before settling in a finally stable state, so they can effectively act as barriers, trapping a nucleus in a local minimum\footnote{These can eventually double beta decay, but all observed half lives of this process are at least a billion times the age of the universe, so this process is negligible}.  

\begin{centering}
\begin{figure*}[h]
\centering
\includegraphics[width=12cm]{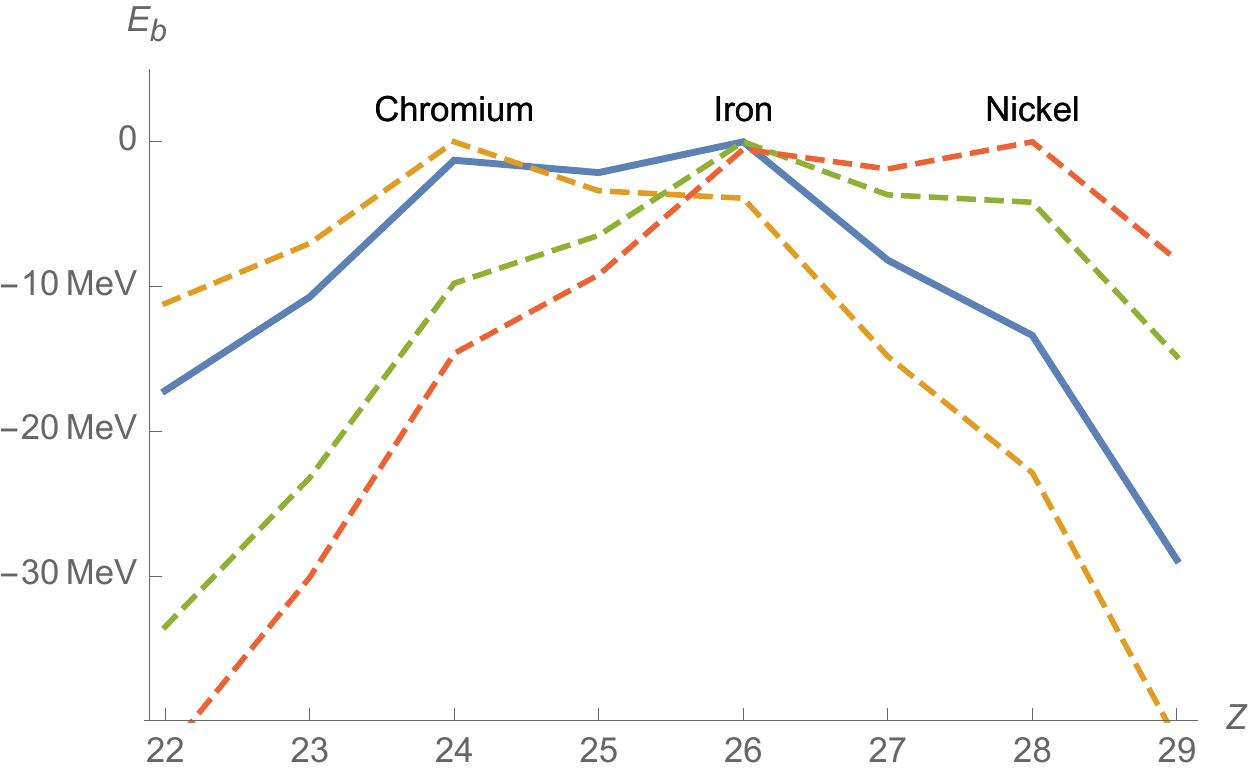}
\caption{The energies of transition metal nuclei for several values of $\alpha$, normalized to the highest binding energy for comparison.  The yellow curve is on the threshold where iron becomes unstable, and will decay ultimately to chromium.  The green and red curves exhibit cases where nickel is stable, and the solid blue line is for the observed value of $\alpha$.}
\label{binding}
\end{figure*}
\end{centering}

We consider that $\beta$ decay will occur if the difference in binding energies exceeds the electron mass.  This tells us nothing about the decay rate, so strictly speaking it ignores the situation where a nucleus may be stable on timescales of the age of the universe, but we neglect this case, as a slight increase in the energy difference would remove this pathology.  It is straightforward to determine the conditions for which the binding energy of iron is the most stable by taking differences
\beq
E_b(56,26)-E_b(56,25)>m_e\quad \implies \alpha<\frac{1}{95}
\eeq
Similarly, nickel-56 will not decay if cobalt-56 has a higher energy, at least in the absence of a catalyst.  The condition for it to decay is
\beq
E_b(56,28)-E_b(56,27)>m_e\quad \implies \alpha>\frac{1}{205}
\eeq
Together, these give a narrow window for which iron is both stable and produced.  This situation is illustrated in Fig. \ref{binding}.  

\section{Iron and Life}\label{life}

Having uncovered a threshold beyond which iron is largely replaced by either nickel or chromium, we must address the issue of how such a replacement would affect the emergence of complex life.  The possible considerations fall into two broad categories: geophysical and biochemical.  We do not pinpoint a clear explanation as to the absolute importance of iron, but outline several plausible explanations, the relative merits of which may be determined after further study.\\

\noindent{\bf Magnetic Field:} Perhaps the most obvious geophysical culprit for iron's importance is Earth's magnetic field.  The majority of Earth's iron has settled to the center to become a partially molten core, where it generates a planetary magnetic field through the dynamo mechanism.  What would the impact be if our core were made out of nickel or chromium instead?  We give evidence that a direct replacement of these elements is unlikely to affect life directly, but suggest an indirect effect that may play a crucial role.  

The magnetic field acts as a shield for high energy cosmic rays, so a first hypothesis may be that for smaller magnetic field, more harmful particles would make it to the surface and potentially be lethal for life.  However, the upper atmosphere would still absorb most of the cosmic rays even in its absence, as outlined in \cite{sagan}.  This is bolstered by the lack of direct correlation between extinction rates and the occasional pole shifts the Earth undergoes, when the field drops to $10\%$ its normal value \cite{gv}.  However, this bombardment would slowly strip particles from the upper atmosphere, and, when acting over geological timescales, would totally deplete it, impacting the biosphere indirectly.

The Earth's magnetic field is generated by the dynamo effect, in which an instability occurs in the magnetohydrodynamics equations for a rotating fluid \cite{stevenson}.  This causes a magnetic field to grow until the Lorentz force on charged particles roughly balances the Coriolis force, creating a magnetic field of strength
\beq
B\sim \sqrt{\frac {\rho\,\Omega}{\sigma}}
\eeq
Corresponding to an Elsasser number of 1.  Here $\rho$ is the density of the fluid, $\sigma$ is the electrical conductivity, and $\Omega$ is the rotation rate.  If the core material is suddenly changed from iron to nickel or chromium, the density would remain practically constant, especially since the number of nucleons remains fixed.  From \cite{egry}, it appears that even the electrical conductivities will not differ by more than a factor of two (Chromium is not tabulated here, but unless it is extremely anomalous compared to the other metals whose electrical properties have been investigated in the molten state, similar conclusions will hold.)  If we take the minimum allowable magnetic field to be ten times less than the present value, this cannot be the cause of inhabitability unless the conductivity changes by two orders of magnitude\footnote{In addition, one may worry that any change may be compensated by an increase in rotation rate.  It appears that an increase by a factor of two will both lead to an inhospitable planet \cite{ybfa} and be highly unlikely to form from the protoplanetary disk \cite{mb}.}.  Finally, we note that the above estimate of the magnetic field is in the interior of the Earth, and the value at the surface scales with the radius of the planet.  As we outline in section \ref{radiolife}, this must be nearly held constant in order for a planet to be habitable.

Thus, if a decreased magnetic field is the reason why iron is singled out in our universe, it cannot be due to the electrical properties of this material directly.  It may be due to planetary history, rather.  It should be noted that chromium, unlike iron and nickel, is a lithophilic element, which means that it binds very readily to rocks present in the crust and mantle.  This could be enough to arrest planetary differentiation, preventing the formation of a well defined core in the first place.\\

\noindent{\bf Biology:} Of the biochemical explanations, there are several reasons why iron may be necessary.  There is the simple fact that iron is one of the essential nutrients required by all life today.  Animals use it to transport oxygen through the bloodstream, plants use it for photosynthesis, and even copying DNA requires iron.  In some parts of the ocean the abundance of iron is the limiting factor determining algae population size, and seeding the region with additional iron fillings produces tremendous algal blooms \cite{buesseler}.  Of course, we cannot deduce that iron is essential for life in general just based off the observation that all life on our planet depends on it, and it is certainly easy to imagine that complex life could in principle arise in its absence, especially from the naive standpoint of a particle physicist that knows very little about biochemistry.  In fact, there are several examples of organisms that do not employ the use of iron the way the rest of life does.  Firstly, the bacteria Borrelia burgdorferi, one of the causes of lyme disease, uses Manganese instead of iron, a trick evolved to overcome the host's usual strategy of turning anemic to starve parasites \cite{lyme}. In addition, even some macroscopic organisms, most notably horseshoe crabs, do not use iron in their blood, but instead rely on copper for oxygen transport.  From these examples, it is certainly plausible that complex life might thrive in an extremely iron poor environment. \\

\noindent{\bf Abiogenesis:}  If iron is not necessary for the existence of life, it may still have been essential for its emergence.  While there is little consensus on the precise details of life's emergence, several leading scenarios either require or are made vastly more plausible with large amounts of iron.  One of the most popular possible explanations is the RNA world hypothesis, in which RNA plays the role of information storage system, enzyme, and expression regulator.  It was recently shown that the presence of iron greatly enhances the catalytic capabilities of RNA \cite{rna}.  An alternative scenario is the iron-sulfur world hypothesis \cite{wachter}, which posits that life first began near iron and sulfur rich deposits near undersea hydrothermal vents.  Thirdly, the process of glycolysis, one of the most ubiquitous methods employed by life to extract energy from sugars, has also recently been shown to be able to occur through the pentose phosphate pathway even in the absence of any enzymes only in ferrous solutions, such as those present in the Archean oceans \cite{ppp}.  In all these cases, iron plays the role of an inorganic cofactor facilitating biotic processes, but it is uncertain whether in its absence another heavy metal could play the same role.  Aside from these, there is a speculative hypothesis that multicellularity was initially driven by magnetobacteria (page 93 of \cite{wb}), that use magnetic field lines to navigate towards favorable conditions, and whose need for a coherent magnetization naturally led to a rigid cell structure, supporting larger sizes and symbiosis.  These bacteria follow magnetic field lines towards regions of higher oxygen content, and naturally clump together once they reach these zones.  Thus, there are several possible reasons for why iron is crucial for the emergence of complex life, but distinguishing among the various different scenarios remains an open question.  Several ways of distinguishing between these are outlined in the conclusions.

\section{Radioactive Elements}\label{radio}

We can place even more restrictive bounds on the fine structure constant if we consider how it influences radioactive elements.  These are present in both the core and the mantle of the Earth, and generate a significant amount of heat, powering the convection that ultimately results in plate tectonics.  This is now recognized as a prerequisite for the development of life on a planet.  It may be surprising at first to hear that life depends crucially on the existence of plate tectonics, especially since its byproducts, earthquakes and volcanoes, rank among the costliest natural disasters that occur.  In fact, by now this is well established geologically \cite{wb,vm}, based off multiple lines of reasoning.  For one, the constant churning through of material, with new mountains rising out of the mantle and being subsequently eroded into the raw materials necessary to assemble life, represents the main source of all of these elements on our crust.  In addition, this weathering acts as a negative feedback loop, which also helps stabilize a habitable range of temperatures over periods of billions of years.  Without plate tectonics, our planet itself would be dead, and as such would be unable to support life.  Slight alterations of the fine structure constant drastically change the half lives of radioactive particles, especially those that undergo alpha decay, and so will have a great influence on this.

The internal heat of the Earth has long been a mystery, ever since the 1840s when Lord Kelvin famously declared that geologists' estimates for the age of the Earth being billions of years was incompatible with the timescale over which such a body would be able to retain its primordial heat, which would have been on the order of millions of years.  Subsequent analysis has overwhelmingly favored the geologists' initial estimates, and has offered two points of contention that Lord Kelvin neglected: that the thermal conductivity of rock under such high temperatures and pressures is much greater, leading to much more heat stored, and that there are radioactive elements throughout the mantle, yielding a continuous supply of further heat (see \cite{england} for a recent review).  Though accurate estimates have been made regarding the relative importance of each of these contributions as far back as \cite{mr}, the Earth's `energy budget' was directly probed only recently.  With the advent of geoneutrino direct detection \cite{geon}, scientists have finally been able to measure that the internal energy budget of the Earth is split roughly evenly between the two.  Of the total $47\pm3$ TW of heat leaving the Earth, KamLAND finds $20\pm9$ TW to be from radioactive decay \cite{kamland}, and Borexino finds $29.5\pm6.5$ TW \cite{borexino}, though these inferences depend on models of how the radioactive elements are distributed throughout the Earth's mantle and crust.  The dominant source of this energy comes from the alpha decay of U238 and Th232 and the beta decay of K40, though the neutrinos generated by potassium decay are not energetic enough to be detected using current methods.  The half life of these isotopes are all in the billion year timeframe (Uranium, in particular, has a half-life $t_{U238}=4.5$Gy, that almost exactly matches the age of the Earth), making it small wonder that they provide the dominant source of energy today:  all nuclides with shorter half lives have long since decayed away, and all nuclides with longer half lives appear practically stable.  

Though at this point there is no way of being certain whether $4.5$ billion years is a typical lifetime for intelligent life to develop, the timescale of billions of years is what is important, since the half lives for alpha decay depend exponentially on the energy released, due to the tunneling nature of the process.  This is encapsulated by the Geiger-Nuttall law \cite{krane}
\beq
\log_{10}(t_{1/2})=\frac{a_0 Z}{\sqrt{\Delta E_b}}-b_0
\eeq
With constants $a_0=1.45$MeV, $b_0=46.83$MeV.  Though a simple phenomenological formula, it typically gives an accuracy of $\sim10\%$, especially for nuclei with even numbers of both protons and neutrons.  

In fact, there is reason to believe that life necessarily takes billions of years to develop intelligence: the similarity to stellar lifetimes has been taken as evidence that this may be the minimum possible time required \cite{anthroevo} (though see \cite{long} for the alternative interpretation that time spent is the circumstellar habitable zone is typically short).  Exactly how much of this time period can be ascribed to the peculiarities of Earth's evolution, and how much is due to a steady growth in complexity, is unimportant, as long as the order of magnitude remains in the billions of years.  Any longer would overstay the lifetime of the sun, and would require orbiting a qualitatively different star.  In addition, it is unknown how the development time of complex life depends on $\alpha$: there is reason to believe that it may, since the chemical timescale does.  This effect would alter the timescales by several hundred million years, a percentage of the timescales we are considering.

Applying the Geiger-Nuttall law to Uranium-235, we can obtain an analytic dependence on $\alpha$ of the decay rate 
\beq
\Delta E_b(\alpha)=E_b(A,Z,\alpha)-E_b(A-4,Z-2,\alpha)-28.3 \text{MeV}+\kappa(\alpha-1/137)
\eeq
This requires the dependence of the He4 binding energy on $\alpha$, but thankfully, this was calculated in \cite{ekllm} to be $\kappa=.61$MeV (though none of our results depend strongly on this value).  If we demand that it is within a factor of 10 of the current age of the Earth, we arrive at the bounds $1/139.4<\alpha<1/137.4$.  In other words, for Uranium to be the dominant decay, $\alpha$ is determined to within $.2\%$!  It is worth remarking that this window in fact lies just outside the value we observe, but this is quite alright, as the accuracy of the Geiger-Nuttall law is actually better than typical in this case.  It would have been possible to set the coefficients to reproduce this single decay rate, or even to use a more refined (still phenomenological) law, as in \cite{gn law}, but we think that the narrowness of this window is sufficient to display the main point, that even a minute variation of the fine structure constant is enough to alter this property of our universe.

\begin{centering}
\begin{figure*}[h]
\centering
\includegraphics[width=12cm]{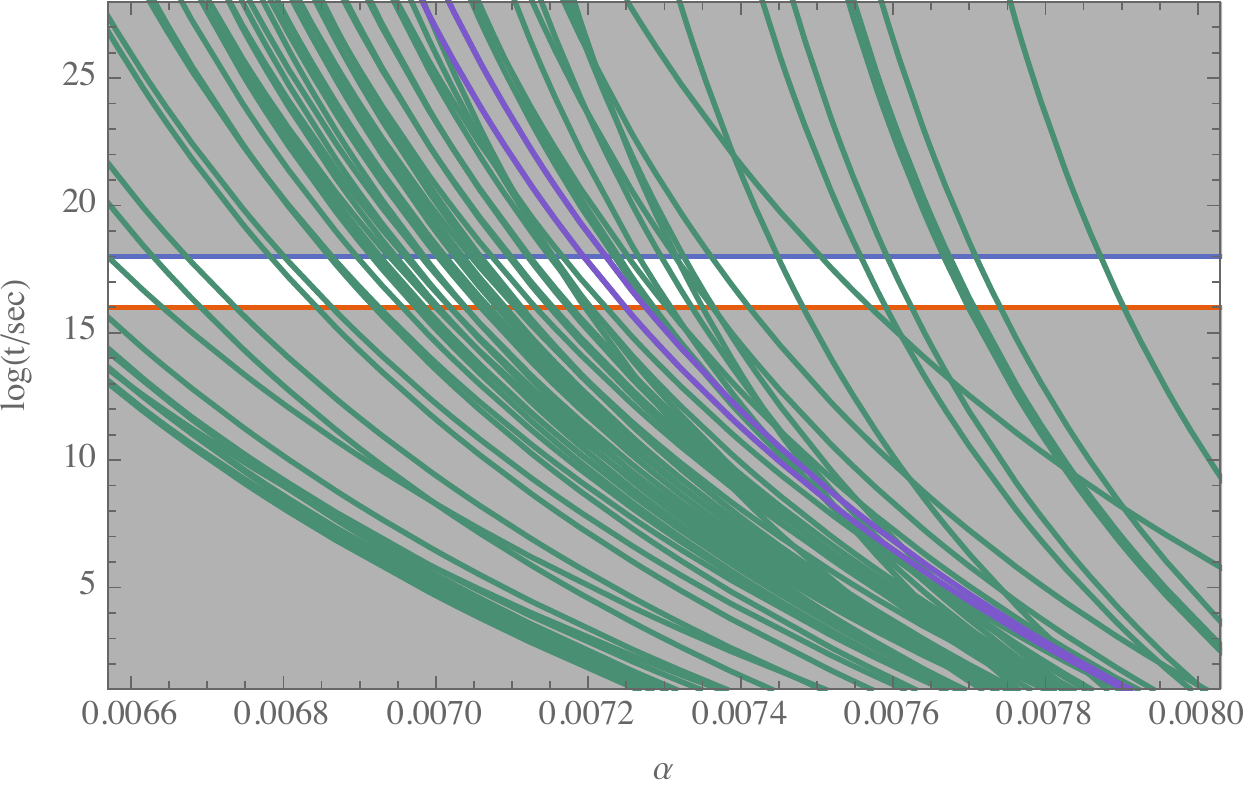}
\caption{Half lives of the various alpha decays observed, as a function of $\alpha$.  The narrow band corresponds to timescales within a factor of 10 of the age of the Earth, as of April 2016.  The curves for U238 and Th232 are highlighted for reference.  It can be seen that if $\alpha$ is changed even slightly, there will be either far more or far fewer decays contributing to the Earth's heat at this time.}
\label{alphadecays}
\end{figure*}
\end{centering}

The reader may object that our reasoning at this point is completely backwards:  in fact there is a large number of radioactive isotopes with half lives ranging from microseconds to quadrillions of years, and U238 is the dominant source of heat because this is the one whose age most closely matches the current age of the Earth.  If we were to vary $\alpha$, U238 would no longer match so perfectly, but surely one of the others would take its place.  In fact this picture is more accurate, as Fig. \ref{alphadecays} indicates.  

This takes the dependence of all 65 alpha decays listed in \cite{w} into account, and shows that actually for the majority of parameter space there is at least one element whose half life lies within an order of magnitude of the age of the Earth.  In fact, if $\alpha$ is allowed to vary over the entire range between which the longest lived and shortest enter the window of Earth's timescale, $82\%$ of the values of $\alpha$ have at least one candidate isotope.  To settle which values are actually allowed we must estimate the heat produced in the Earth's mantle, not just the presence of such isotopes.  We use
\beq
Q(t)=\sum_i \frac{\rho_i\,V_{\text{Earth}} \Delta E_i}{t_i\, 2^{t/t_i}}
\eeq
Where $\rho_i$ is the cosmic abundance of a given isotope, $V_{\text{Earth}}$ is the volume of the Earth's mantle, $\Delta E_i$ is the energy liberated by decay, and $t_i$ its half life.  The sum is over all species.  We have explicit expressions for the dependence of the energies and half lives on $\alpha$, but the densities are harder to come by.  All of these elements are produced in the r-process, as evidenced by the fact that they are all along the neutron-drip line, which indicates that they were produced by neutron accretion much faster than they could beta decay to the more preferred ratio.  The values of the abundances are dependent on the details of this process, and so we consult the relative abundances quoted in \cite{ctt}.  These do not take into account any dependence on $\alpha$, but we do not expect the details to strongly depend on its value, given the nature of isotope formation is due to neutron accretion cross sections.  In any event, we discount the possibility that the change in abundance would somehow conspire to keep the total heat generated at this particular moment in time constant, so we are confident that using these values produces a reasonable estimate.

This calculation has been undertaken, and is reported in Fig. \ref{heatboth} below.  The left plot shows the heat generated by radioactive elements as a function of time, for several different values of the fine structure constant.  The rates resemble a series of plateaus, as supplies of short lived nuclei are successively exhausted, so that the value at any given moment is always dominated by the isotope with the decay rate closest to that time.  From the curves it can be seen that increasing $\alpha$ causes alpha decay to be more efficient, leading to the scenarios with a rapid initial release of energy, which nevertheless is practically depleted by our point in time, $\log_{10}(t_{\text{Earth}}/\text{sec})=17.1564$.  Likewise, decreasing $\alpha$ by just $10\%$ causes the rates to be much slower, reserving the nuclear energy to be doled out over a much longer period of time.  

\begin{centering}
\begin{figure*}[h]
\centering
\includegraphics[width=16cm]{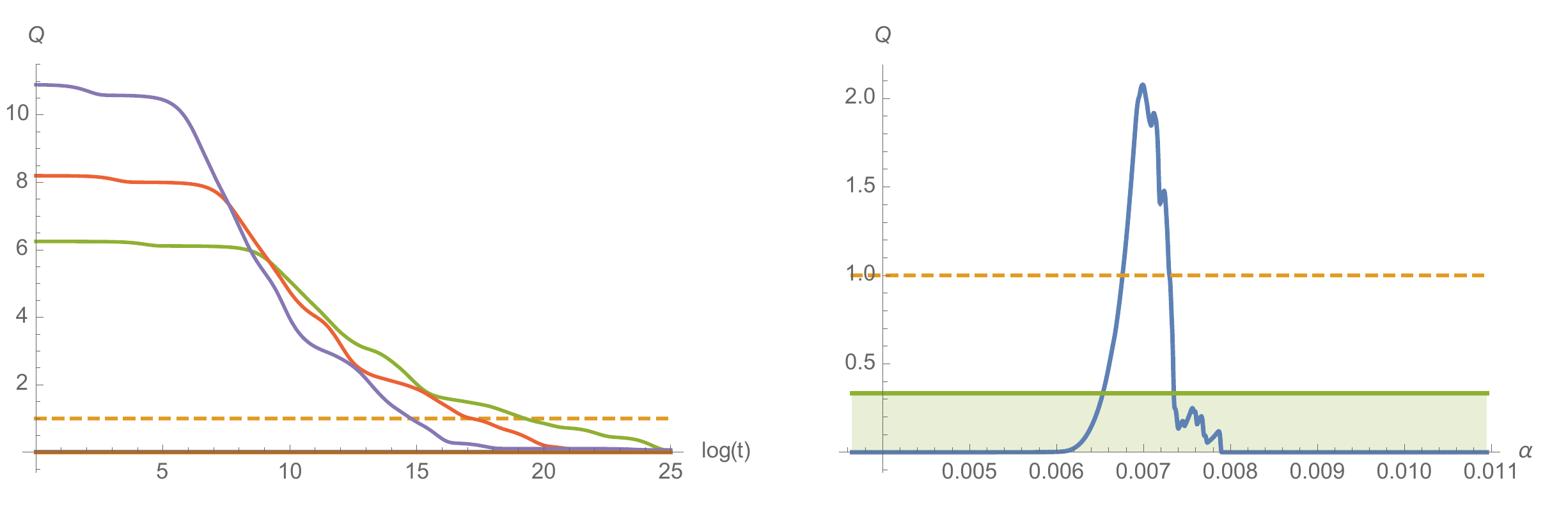}
\caption{{\bf Left}: Radiogenic heat release as a function of time, normalized to today's value in our universe, for several different values of $\alpha$, ranging from $.99\alpha_0$ (highest initial heat) to $1.01\alpha_0$ (lowest initial heat).  The differences of just a percent level change can be seen to be very drastic.  {\bf Right}:  The heat released at the current age of the Earth as a function of $\alpha$.  The bound $Q/3$, at which point plate tectonics comes to a halt, is also shown.}
\label{heatboth}
\end{figure*}
\end{centering}

Fig. \ref{heatboth}b shows the amount of energy released at our moment in time as a function of $\alpha$.  From here the strong dependence can readily be seen.  Decreasing $\alpha$ from its observed value leads to an increase in the amount of radiation that would occur, until it peaks at a value around 2.6 times the amount we observe.  This corresponds to the region evident in Fig \ref{alphadecays} with a great number of active decays on this timescale.  Outside this region the rate quickly drops, due to the absence of suitable decays.  Below, we argue that a value of $Q/3$ would be inhospitable based off the analysis done in \cite{vos}, which bounds the fine structure constant to be within
\beq
\frac{1}{153}<\alpha<\frac{1}{136}
\eeq

We have tested the robustness of this conclusion to the imprecision of the Geiger-Nuttall law by artificially changing the half lives by random values of order $10\%$, and found that the spread of the upper bound is $133\pm4$, while the lower bound is $152.9\pm.5$.

Before we enumerate the various caveats and assumptions in this analysis, let us note that there are still around 100 beta decays that may potentially become relevant for different values of $\alpha$, but we argue that they are in fact usually not:  beta decays are governed by a weak interaction induced species change, and as such depend on the binding energy only polynomially, not exponentially, as $t\sim 1/(G_F^2m_e^2 E_b^3)$ \cite{opqccv} for `forbidden' transitions, and even $t\sim 1/(G_F^2m_e^5)$ for the `superallowed' transitions \cite{krane}.  The typical lifetimes have a very large spread, but they do not depend on $\alpha$ anywhere near as drastically as the alpha decays, except in the region where $E_b\approx0$.  For this region, however, the amount of heat generated by these decays would be negligible, and so we are justified in neglecting them.  


A few selection effects that can possibly affect the conclusion we have drawn: In the chart \cite{w}, some decays that are reported as in the beta channel may become dominated by alpha for different values of the fine structure constant, and vice-versa.  However, this should not be the case, because the ratio $A/Z$ in the set ranges from $2.33$ to $2.78$, meaning that every sample is very neutron-rich.  Such elements will always have a very fast beta decay, on the order of seconds, so any crossovers will not be relevant to billion year timescales.  A bigger worry is that the relevant decays for drastic values of $\alpha$ would have pathologically long or short decays in our universe, leaving them unreported in the table we used.  However, these effects will not contribute significantly to our conclusions either: we note the general trend that extremely short lived nuclei tend to have higher $Z$.  This means that any half-life that is greatly increased in other universes corresponds to an element with very low abundance.  Long lived nuclei, with smaller $Z$, also tend to be closer to the stable ratio, meaning that, while their abundances will be larger, their main decay channels will remain beta decay.  Since we have argued that these will not be as drastically altered, these decays will remain close to their values in our universe.

\section{Radioactivity and Life}\label{radiolife}

{\bf Tectonics:} The feature that must be maintained is the fact that the crust of the Earth has plate tectonics.  This represents an intermediate regime between a crust that is entirely molten and a ``stagnant lid'', where no subduction occurs.  The actual conditions for plate tectonics are still in the process of being established, and there are conflicting reports in recent literature as to when this transition actually occurs.  The basic condition for plate tectonics is to have the convective stresses of the mantle roughly equal the yield stress of the lithosphere, which in turn causes it to fail.  The subtleties lie in determining the dependence of these quantities on multiple variables, such as the depth of the core-mantle boundary and brittle-ductile transitions, the density, thermal conductivity and Rayleigh number of the mantle, the internal heat generated, the relative abundances of various minerals, and, perhaps most importantly, the amount of water present both on the surface and in the interior \cite{allyall1,allyall2,allyall3,allyall4}.  Nevertheless, \cite{vos} explicitly studied the effects of internal heat generation on plate formation, keeping all other parameters fixed, and concluded that if the internal heat were $1/3$ its current value, then the lid would be thicker, halting activity.  We can take this to be our lower bound.  We conservatively assume that the peak value of the heat production would still lead to plate tectonics, though this amount of radioactivity has likely never occurred on our planet.\\

\noindent{\bf Mass of the Earth:} We have required the planets with plate tectonics to be of Earth-like size, which to this point has not been justified.  If there is not enough heat in a planet like Earth, we may still expect that larger planets would have the necessary amount of radioisotopes to support plate tectonics.  Likewise, if Earth were enriched enough to become molten, a smaller planet may have plates.  However, there is the further requirement that restricts the size of habitable planets: they must maintain a suitable atmosphere.  Atmospheres are composed of a variety of gases, but for planets of Earth-like radii with habitable temperatures, the average speed of gas particles is comparable to the escape velocity of the planet.  This allows the lighter species like hydrogen and helium to escape, preventing the runaway process of gas accretion that would lead to Neptune-like type planets with intense pressures and winds, but retains the heavier components of the atmosphere, preventing our planet from becoming Mars-like with essentially no atmosphere.  

The condition for a gas to be marginally bound can be found by equating the thermal energy to the kinetic energy at escape velocity, $3kT/2=\frac{4\pi}{3}G\rho R^2 m$.  If the thermal energy is any greater, the particle will evaporate, and if it's any less it will remain captured\footnote{In actuality, exact equality must be replaced with the condition that flux must be negligible on the timescale of a billion years \cite{pl}.}.  From here, we can deduce the $\alpha$ dependence by noting that the density scales as $\alpha^3$, as the volume of a unit cell for a solid is proportional to the Bohr radius cubed.  The temperature is set by the Rydberg, the energy scale of atomic interactions, which is proportional to $\alpha^2$.  In addition, dependence on the other constants of nature can be introduced to this expression, but here we remain focused on $\alpha$ for simplicity.  This restricts the radius of the planet to the very narrow range
\beq
.7<\sqrt{\frac{\alpha}{\alpha_0}}\frac{R}{R_{\text{Earth}}}<1.6
\eeq
The upper bound is in excellent agreement with recent observations of exoplanet populations \cite{rogers}.  However, the variation of the upper and lower bounds with our changes in $\alpha$ do not vary more significantly than the intrinsic scatter of observed planet populations, and so should be unimportant.  In other words, there are two habitability criteria, atmospheric and tectonic, that must line up for a universe to be suitable for life.  Without the levels of radioactivity we observe, these two windows would not align in our universe, but, thanks to the Earth's radioactive interior, they luckily do.\\

\noindent{\bf Metallicity:}  One important point is the abundance of heavy elements itself:  how environmental is this in our solar system, and could any change in heat generation be offset by a change in the overall abundance?  We believe that the answer is no, for several reasons.  First, note that the range of observed metallicities only spans about an order of magnitude \cite{disk}, with the center of the galaxy being richer and the outskirts lacking.  This limits the offset that can be provided, especially since the continued injection of short-lived radionuclides \cite{53} indicates that the time it took our protoplanetary disk to collapse was longer than the time it took our young solar system to traverse its parent cluster.  This implies that the solar system uniformly sampled its environment by the time planets formed, reducing possible environmental variance\cite{al}.  In addition, it was recently shown \cite{fv} that stars with higher metallicity may correlate with having a hot Jupiter, whose inward migration would have destabilized Earth's orbit early on in its history.  (However, see \cite{rub} for the counterpoint that Earth-like planets may commonly form after this migration has taken place.)

This is not to say that it is impossible to compensate for a highly radioactive universe by simply dialing another parameter to increase the metallicity, but simply that within a single universe we would still expect the metallicity of different solar systems to not vary substantially.  How to achieve such a compensation, if possible, remains to be shown, and would not diminish the importance of the anthropic boundary we have found here.  In fact, it would not even alter the dimensionality of the allowable subspace of parameter space required for our considerations, but at most would make the subspace tilted or curved.\\ 


\noindent{\bf Differentiation:}  One simplification of our analysis is the assumption that the chemical abundances in Earth's interior are identical to their cosmic abundances.  In reality, the two will be slightly different, as refractory elements, which have a high condensation temperature, will preferentially clump during planet formation, leaving the volatile ones to be blown away by solar winds.  The extent of this differentiation is not currently completely known, as it depends on many of the details of planet formation and the early solar system that we do not yet have access to \cite{ms}.  However, it is beyond the scope of this paper to factor current models of the Earth's bulk composition to this analysis.  \\

\noindent{\bf R Process:}  A further simplification in this analysis was that in creating the abundance estimates we have only taken into account data about the nuclear number $A$, and not $Z$.  We have tacitly assumed that for a given number of nucleons, the relative abundances are of similar magnitude over the $A\sim2.5Z$ region, but this is not necessarily the case.  These numbers are only reported for the most important isotopes \cite{ctt2}, and even then there are uncertainties on the order of at least $10\%$.  But, of the numbers that were quoted, all are essentially $O(1)$, which bolsters our confidence that our estimate will be relatively good.  It is much more difficult to calculate the expected abundances from first principles, since this requires a table of the nuclear cross sections, which have inherent uncertainties, as well as the temperature and density of the environment where the r process takes place, which is poorly understood \cite{ctt,iliad}.  Because of this, we are forced to delay a more thorough calculation until the time when we have more information on the r process site.

\section{GUTs}\label{nitty}

One other scenario that would determine the value of the fine structure constant is grand unified theories (GUTs), and indeed this reasoning has been successfully applied to constrain its value to within a very narrow window around what we observe \cite{bt}.  The logic is as follows:  The strength of the weak and electromagnetic forces increase with energy due to quantum effects, and the strong force decreases, and appear (with some additional ingredients, such as supersymmetry) to converge to the same value at the energy scale $\sim10^{16}$GeV, where the three forces can be unified into a larger gauge group (see \cite{raby} for a review).  If the fine structure constant had been lower, then, this unification would have occurred at a higher energy.  If $\alpha<1/180$ this energy would have been pushed to above the Planck scale, where we know our current description of physics breaks down, and unification would be deemed highly suspect.  Likewise, GUTs typically predict that the proton is not absolutely stable, with a decay rate that depends on the strength of the coupling.  If $\alpha>1/85$, protons would decay faster than the lifetime of typical stars, and so we arrive at a natural window reliant on the existence of unification.  This is compared to the iron peak and radioactivity windows in Fig. \ref{window} in the introduction.

We see that all three considerations require that the fine structure constant be in the vicinity of the observed value.  These GUT bounds have a few caveats, however.  Firstly, the generic prediction of proton decay is considered a major downfall of the simplest GUT models, but is very dependent on the actual details of the scenario, which may effectively alleviate the upper bound for specific proposals.  Secondly, the lower bound depends on the particle content, though the running is dominated by low mass particles, as they cause the constant to run for longer periods than those that decouple at very high energies.  Thus, we treat the GUT window as somewhat malleable.

A greater puzzle emerges, though:  if there are two requirements on $\alpha$, one as a demand on the underlying theory, and one for the emergence of life, how is it that they both just happen to overlap?  The simplest interpretation is that they cannot both be true, and that one would have to be a mirage.  In this scenario, if strong evidence is found for GUTs, then we would conclude that life does not depend strongly on iron or plate tectonics, and may even expect to find complex life in regions of the universe where these are absent.  Conversely, if we manage to prove that complex life is impossible without these, then we may not expect that the other window be valid, i.e. that the GUT hypothesis is false.  

However, it is too strong to turn these scenarios into a strict dichotomy.  It may instead suggest that there is some freedom in the GUT window that can be chosen by demanding that it overlap with the habitability windows.  Certainly, there are ways of altering the position of the window, such as changing the number of light generations, the scale of supersymmetry breaking, or the vacuum expectation value of the field that breaks the unified symmetry.  The overlap of these windows may be taken as an indication that one should not be expecting a GUT with a very rigid structure, but would instead expect to find that the theory has at least one tunable parameter, giving it the freedom to automatically satisfy this requirement for some value.  This demonstrates that further considerations of suitability for life may dictate the types of high energy physics to expect in future experiments.

\section{Conclusions and Future Tests}\label{conc}

We have argued that a small change in the fine structure constant would produce drastic changes in some of the properties of rocky planets that would greatly affect the habitability of these worlds.  This comes about in two separate ways: small variations, around $6\%$, would alter nuclear decay rates to the point where the interior of the Earth would contain significantly less radioactive isotopes today.  Varying by $34\%$ results in a universe with drastically less iron. Using the principle of living dangerously, we conclude that since our universe is so sensitive in these ways, that these two features must play an important role in the development of complex life.  We have only alluded to several possible reasons for the case of iron, and now turn to discussing how to differentiate between the various explanations put forth.  This current ambiguity can actually be viewed as a benefit of the proposed explanation, for as it stands now the dependence of life on iron counts as a prediction that can be tested and falsified.

For the importance of iron, we have put forward two plausible scenarios, the first being that it provides the magnetic shield necessary to protect the atmosphere, oceans, and life.  We found that the properties of molten iron are not very different from those of chromium or nickel, so we concluded that if this explanation is to be correct, there must be some difference in the geological history of at least one of these other elements that would prevent a core from forming in the first place, precluding the dynamo mechanism from being operational.  This remains a conjecture at this stage, but any indication of this would lend strong support to this interpretation of the boundary.  

Our second suggestion is that biological processes depend on iron in some crucial way.  Though there are a number of examples in which specific organisms have foregone some of the most common uses of iron, there are several different indications that iron may have been crucial in the process of generating life initially.  It may soon be within the realm of experiment to recreate the necessary conditions to generate such extraordinary complexity, and once this is achievable, it will be within our capabilities to use the amount of iron as a control parameter.  Additionally, we are on the verge of directly sensing the atmospheres of large numbers of distant exoplanets, with upcoming space missions the James Webb Space Telescope \cite{jwst}, CHEOPS \cite{cheops} and TESS \cite{tess}.  Searching for signatures of microbial life on totally nonmagnetized, or totally anemic, planets would provide useful input on this issue.  In other words, we have a sense that life is inescapably ironic, though we are not yet exactly sure how.

A further requirement we have made is to demand the existence of plate tectonics, which is crucially dependent on the presence of sufficient interior heat.  Without such heat, the radii required for tectonics and that required for a suitable atmosphere would not coincide, and no planet would be suitable for complex life as we know it.  Thus, $\alpha$ emerged as a tunable parameter, that must have been set to the value necessary for this coincidence as a selection effect.  However, since the understanding of plate tectonics is still in its relative infancy, the possibility remains that the requirements for its operation be strongly dependent on some parameter that can easily vary within our single universe, such as the ratio of certain minerals or the amount of water present.  Continued progress in this field will be able to inform us of how uniform the tectonic activity of disparate stellar systems actually is, or whether the existence of plate tectonics is, should we say, set in stone.

There is also an apparent coincidence that the habitability window overlaps with the GUT window, which may imply that we should expect a grand unified theory with enough freedom to encompass this without fortuity.  Alternatively, one may not assign very high significance to this coincidence at all:  If we naively ask what the probability that these two windows overlap is, just based off the sizes of the two windows assuming they were uniformly distributed over 0 and 1, we conclude that the chances are about 1 in 142.  This is a coincidence, truly, but hardly one that can be employed to strongly disfavor a whole paradigm.  Perhaps what we are forced to conclude is that in universes with iron (and plate tectonics), life is at least 142 times more likely to evolve than in universes without.  This may sound like offhand flippancy, but it actually may be turned into a useful test of this idea: if we imagine in the very distant future, when our descendants have a thorough catalog of the places life emerged, if they find that its nucleation only occurs about 100 times more frequently in the presence of iron or plates, we can chalk the overlap with the GUT window up to a coincidence, and continue searching for signs of a rigid GUT.  If, on the other hand, life is found to be extraordinarily more likely in these habitable places, we would expect flexible GUTs, or none at all.  All this is to say, we will eventually find guts, but whether it will be on distant planets or in future colliders remains to be seen.

Though most of the results in this paper have been phrased in terms of $\alpha$ alone, it is important to note that they will depend on other parameters as well.  This will be important if these results are to be incorporated into a broader context.  To date, there has been no systematic enumeration of the dependence of all the various anthropic constraints on the fundamental physical parameters.  An interesting question is that, given the parameter space of several dozen that completely specify the microphysical properties and initial conditions of a universe, what subspace within this is compatible with complex life?  What should we expect the dimensionality of this subspace to be?  For every new requirement we find, the dimension of this subspace is reduced by one, even if we do not explicitly track the dependence of all parameters in the initial uncovery (excluding any accidental correlations with other constraints).  Since we have found two different constraints in this paper, we may conclude that this will ultimately reduce the dimensionality of this hypersurface by two.

We hope to have conveyed that there are additional anthropic boundaries awaiting consideration, that the time is ripe to incorporate results from planetary science into these considerations, and that the information we will learn in the coming decades will shed light on the structure of both fundamental physics and the multiverse.\\

{\bf \noindent Acknowledgements}

\smallskip
I would like to thank Richard Boyle and Martin Sloth for useful comments and discussions.  The CP$^3$-Origins center is partially funded by the Danish National Research Foundation, grant number DNRF90.


\begin{thebibliography}{99}

\bibitem{just so} 
  C.~J.~Hogan,
  ``Why the universe is just so,''
  Rev.\ Mod.\ Phys.\  {\bf 72}, 1149 (2000)
  doi:10.1103/RevModPhys.72.1149
  [astro-ph/9909295].

\bibitem{earf}
S. K. Runcorn (ed.),
``The Application of Modern Physics to The Earth and Planetary Interiors",
Wiley-Interscience, London, 1969
SBN: 471 74505 7.

\bibitem{numbers}
P. A. M. Dirac,
``The Cosmological Constants,"
Nature {\bf 139}, 323 (1937)
doi:10.1038/139323a0.

\bibitem{dicke1}
R. H. Dicke,
{\it The Theoretical Significance of Experimental Relativity},
Gordon and Breach, New York, 1964,
ISBN-13: 978-0677002200.

\bibitem{dicke2}
R. H. Dicke, 
``Dirac's cosmology and Mach's principle." 
Nature {\bf 192} 440 (1961)
doi:10.1038/192440a0

\bibitem{coin} 
  C.~H.~Lineweaver and C.~A.~Egan,
  ``The Cosmic Coincidence as a Temporal Selection Effect Produced by the Age Distribution of Terrestrial Planets in the Universe,''
  Astrophys.\ J.\  {\bf 671}, 853 (2007)
  doi:10.1086/522197
  [astro-ph/0703429 [ASTRO-PH]].

\bibitem{standp} 
  F.~C.~Adams,
  ``Constraints on Alternate Universes: Stars and habitable planets with different fundamental constants,''
  JCAP {\bf 1602}, no. 02, 042 (2016)
  doi:10.1088/1475-7516/2016/02/042
  [arXiv:1511.06958 [astro-ph.CO]].
  
\bibitem{bt}
J. Barrow and F. Tipler,
{\it The Anthropic Cosmological Principle},
Oxford, Oxford, 1988,
ISBN-13: 978-0192821478.

\bibitem{ksas}
R. A. King, A. Siddiq, W. D. Allen and H. F. Schaefer,
``Chemistry as a function of the fine-structure constant and the electron-proton mass ratio",
Phys. Rev. A {\bf 81}, 042523 (2010)
doi:10.1103/PhysRevA.81.042523.

\bibitem{hn} 
  L.~J.~Hall and Y.~Nomura,
  ``Evidence for the Multiverse in the Standard Model and Beyond,''
  Phys.\ Rev.\ D {\bf 78}, 035001 (2008)
  doi:10.1103/PhysRevD.78.035001
  [arXiv:0712.2454 [hep-ph]].
  
\bibitem{schellekens} 
A.~N.~Schellekens,
``Life at the Interface of Particle Physics and String Theory,''
Rev.\ Mod.\ Phys.\  {\bf 85}, no. 4, 1491 (2013)
doi:10.1103/RevModPhys.85.1491
[arXiv:1306.5083 [hep-ph]].

\bibitem{tr} 
M.~Tegmark and M.~J.~Rees,
``Why is the Cosmic Microwave Background fluctuation level 10**(-5)?,''
Astrophys.\ J.\  {\bf 499}, 526 (1998)
doi:10.1086/305673
[astro-ph/9709058].

\bibitem{page} 
  D.~N.~Page,
  ``Anthropic estimates of the charge and mass of the proton,''
  Phys.\ Lett.\ B {\bf 675}, 398 (2009)
  doi:10.1016/j.physletb.2009.04.021
  [hep-th/0302051].

\bibitem{pl} 
  W.~H.~Press and A.~P.~Lightman,
  ``Dependence Of Macrophysical Phenomena On The Values Of The Fundamental Constants,''
  Submitted to: Phil. Trans. Roy. Soc. (Lond) A.

\bibitem{ocs} 
  H.~Oberhummer, A.~Csoto and H.~Schlattl,
  ``Stellar production rates of carbon and its abundance in the universe,''
  Science {\bf 289}, 88 (2000)
  doi:10.1126/science.289.5476.88
  [astro-ph/0007178].

  \bibitem{ekllm} 
  E.~Epelbaum {\it et al.},
  ``Dependence of the triple-alpha process on the fundamental constants of nature,''
  Eur.\ Phys.\ J.\ A {\bf 49}, 82 (2013)
  doi:10.1140/epja/i2013-13082-y
  [arXiv:1303.4856 [nucl-th]].

\bibitem{rees} 
  B.~J.~Carr and M.~J.~Rees,
  ``The anthropic principle and the structure of the physical world,''
  Nature {\bf 278}, 605 (1979).
  doi:10.1038/278605a0
  
  M.~J.~Rees,
  ``Cosmology and the multiverse,''
  In B. Carr (ed.): {\it Universe or multiverse} 57-75 

\bibitem{dipins} 
  L.~A.~Barnes,
  ``Binding the Diproton in Stars: Anthropic Limits on the Strength of Gravity,''
  JCAP {\bf 1512}, no. 12, 050 (2015)
  doi:10.1088/1475-7516/2015/12/050
  [arXiv:1512.06090 [astro-ph.SR]].

\bibitem{bradford}
R. A. W. Bradford,
``The effect of hypothetical diproton stability on the universe'',
J. Astrophys. and Astron. {\bf 30}, 119 (2009)
doi:10.1007/s12036-009-0005-x.

\bibitem{df} 
  T.~Dent and M.~Fairbairn,
  ``Time varying coupling strengths, nuclear forces and unification,''
  Nucl.\ Phys.\ B {\bf 653}, 256 (2003)
  doi:10.1016/S0550-3213(03)00043-9
  [hep-ph/0112279].

\bibitem{dinter} 
  S.~Dinter {\it et al.} [ETM Collaboration],
  ``Sigma terms and strangeness content of the nucleon with $N_f=2+1+1$ twisted mass fermions,''
  JHEP {\bf 1208}, 037 (2012)
  doi:10.1007/JHEP08(2012)037
  [arXiv:1202.1480 [hep-lat]].

\bibitem{iliad}
C. Iliadis,
{\it Nuclear Physics of Stars},
Wiley, Weinheim, 2007,
ISBN 978-3-527-61876-7.

\bibitem{krane}
K. Krane,
{\it Introductory Nuclear Physics, 3rd Ed.},
Wiley, Singapore,1988,
ISBN 978-0-471-80553-3.

\bibitem{sagan}
C. Sagan,
``Is the Early Evolution of Life Related to the Development of the Earth's Core?"
Nature {\bf 206}, 448 (1965)
doi:10.1038/206448a0.

\bibitem{gv}
K. H. Glassmeier, and J. Vogt,
``Magnetic Polarity Transitions and Biospheric Effects. Historical Perspective and Current Developments",
Space Sci. Rev. {\bf 155}, 387 (2010)
doi:10.1007/s11214-010-9659-6.

\bibitem{stevenson}
D. J. Stevenson, 
``Planetary magnetic fields",
Earth and Planetary Science Letters, {\bf 208} (2003)
http://dx.doi.org/10.1016/S0012-821X(02)01126-3.
    
\bibitem{egry}
I. Egry, 
``Structure and Properties of Molten Metals", 
In S. Seetharaman (ed.), {\it Treatise on Process Metallurgy}, Elsevier, Boston, 2014, Pages 61-148, 
ISBN 9780080969862, 
http://dx.doi.org/10.1016/B978-0-08-096986-2.00007-2.

\bibitem{ybfa}
J. Yang, G. Bou\'e, D. C. Fabrycky, and D. S. Abbot,
``Strong Dependence of the Inner Edge of the Habitable Zone on Planetary Rotation Rate",
Astrophys. J. Lett. {\bf 787}, L2 (2014)
doi:10.1088/2041-8205/787/1/L2
[arXiv:1404.4992[astro-ph.EP]].

\bibitem{mb}
Y. Miguel, and A. Brunini,
``Planet formation: statistics of spin rates and obliquities of extrasolar planets",
Month. Not. RAS {\bf 406}, 1935 (2010)
 doi = {10.1111/j.1365-2966.2010.16804.x}
 [arxiv:1004.1406[astro-ph.EP]].

\bibitem{buesseler}
K. O. Buesseler,
``Fertilizing the ocean with iron",
Ann. Rep. Woods Hole Oceanograph. Inst. (1999)

\bibitem{lyme}
J. D. Aguirre {\it et al.},
``A Manganese-Rich Environment Supports Superoxide Dismutase Activity in a Lyme Disease Pathogen, Borrelia burgdorferi"
J. Biol. Chem. jbc.M112.433540. (2013) 
doi:10.1074/jbc.M112.433540.

\bibitem{rna}
C. Hsiao {\it et al.},
``RNA with iron(II) as a cofactor catalyses electron transfer"
Nature Chem. {\bf 5}, 525 (2013) 
doi:10.1038/nchem.1649.

\bibitem{wachter}
G. W\"{a}chtersh\"{a}user,
``The Case for a Hyperthermophilic, Chemolithoautotrophic Origin of Life in an Iron-Sulfur World",
In J. Wiegel and M. W. W. Adams (ed.), {\it Thermophiles: The Keys to Molecular Evolution and the Origin of Life?}, Taylor $\&$ Francis, London, 1998, Pages 47-58,
ISBN 0-203-48420-7.

\bibitem{ppp}
M. A. Keller, A. V. Turchyn, and M. Ralser, 
``Non-enzymatic glycolysis and pentose phosphate pathway-like reactions in a plausible Archean ocean''. 
Molecular Systems Biology {\bf 10(4)}, 725 (2014) 
doi:10.1002/msb.20145228.

\bibitem{wb}
P. D. Ward, D. Brownlee,
{\it Rare Earth: Why Complex Life Is Uncommon in the Universe},
Copernicus Books, New York, 2003,
ISBN 0-387-95289-6.

\bibitem{vm}
J. W. Valentine, and E. M. Moores. 
``Global Tectonics and the Fossil Record". 
The Journal of Geology {\bf 80.2} (1972): 167

J. W. Valentine and E. M. Moores,
``Plate-tectonic Regulation of Faunal Diversity and Sea Level: a Model'',
Nature {\bf 228}, 657 (1970)
doi:10.1038/228657a0.

\bibitem{england}
P. England,
``John Perry's neglected critique of Kelvin's age for the Earth: A missed opportunity in geodynamics",
GSA Today, {\bf 17}, (2007)
doi: 10.1130/GSAT01701A.1.

\bibitem{mr}
D. McKenzie, F. M. Richter,
``Parameterized thermal convection in a layered region and the thermal history of the Earth",
J. Geophys. Res. {\bf B12}, 11667 (1981)
doi:10.1029/JB086iB12p11667. 

\bibitem{geon} 
  G.~Bellini, A.~Ianni, L.~Ludhova, F.~Mantovani and W.~F.~McDonough,
  ``Geo-neutrinos,''
  Prog.\ Part.\ Nucl.\ Phys.\  {\bf 73}, 1 (2013)
  doi:10.1016/j.ppnp.2013.07.001
  [arXiv:1310.3732 [physics.geo-ph]].

\bibitem{kamland}
The KamLAND Collaboration,
``Partial radiogenic heat model for Earth revealed by geoneutrino measurements",
Nature Geosci. {\bf 4}, 647 (2011)
doi:10.1038/ngeo1205.

\bibitem{borexino} 
  R.~Roncin {\it et al.} [Borexino Collaboration],
  ``Geo-neutrino results with Borexino,''
  J.\ Phys.\ Conf.\ Ser.\  {\bf 675}, no. 1, 012029 (2016).
  doi:10.1088/1742-6596/675/1/012029

\bibitem{anthroevo}
B. Carter,
``The anthropic principle and its implications for biological evolution?,
Phil. Trans. Roy. Soc. Lond. {\bf A310} (1983) 347.

\bibitem{long}
F. Simpson,
``The longevity of habitable planets and the development of intelligent life",
 [arXiv:1601.05063[astro-ph.EP]].

\bibitem{gn law}
C. Qi {\it et al.},
``On the validity of the Geiger-Nuttall alpha-decay law and its microscopic basis",
Phys. Lett. B, {\bf 734}, 203
http://dx.doi.org/10.1016/j.physletb.2014.05.066.

\bibitem{w}
``List of radioactive isotopes by half-life." Wikipedia: The Free Encyclopedia. Wikimedia Foundation, Inc. 22 July 2004. Web. 2 February 2016.
 
 \bibitem{ctt}
 J. J. Cowan, F. K. Thielemannm, and J. W. Truran,
 ``Nuclear chronometers from the r-process and the age of the galaxy",
 Astrophys. J. {\bf 323}, 543 (1987)
 doi:10.1086/165850.

\bibitem{vos} 
  D.~Valencia, R.~J.~O'Connell and D.~D.~Sasselov,
  ``Inevitability of Plate Tectonics on Super-Earths,''
  Astrophys.\ J.\  {\bf 670}, L45 (2007)
  doi:10.1086/524012.

\bibitem{opqccv} 
  K.~A.~Olive, M.~Pospelov, Y.~Z.~Qian, A.~Coc, M.~Casse and E.~Vangioni-Flam,
  ``Constraints on the variations of the fundamental couplings,''
  Phys.\ Rev.\ D {\bf 66}, 045022 (2002)
  doi:10.1103/PhysRevD.66.045022
  [hep-ph/0205269].

\bibitem{allyall1}
C. O'Neill, A.M. Jellinek, and A. Lenardic,
``Conditions for the onset of plate tectonics on terrestrial planets and moons'',
Earth and Planetary Sci. Lett. {\bf 261}, 20 (2007) 
http://dx.doi.org/10.1016/j.epsl.2007.05.038.

\bibitem{allyall2}
C. O'Neill, and A. Lenardic,
``Geological consequences of super-sized Earths",
Geophys. Res. Lett. {\bf 34}, L19204 (2007)
doi:10.1029/2007GL030598. 
 
 \bibitem{allyall3}
 B. J. Foley, D. Bercovici, and W. Landuyt,
``The conditions for plate tectonics on super-Earths: Inferences from convection models with damage'',
Earth and Planetary Sci. Lett. {\bf 331}, 281 (2012) 
http://dx.doi.org/10.1016/j.epsl.2012.03.028.
 
 \bibitem{allyall4}
 Y. Alibert,
``On the radius of habitable planets",
Astron. and Astrophys. {\bf 561}, A41 (2014)
doi:10.1051/0004-6361/201322293
[arXiv:1311.3039[astro-ph.EP]].

\bibitem{rogers}
L. A. Rogers,
``Most 1.6 earth-radius planets are not rocky",
The Astrophys. J. {\bf 801}, 1 (2015)
doi:10.1088/0004-637X/801/1/41
[arXiv:1407.4457 [astro-ph]].

\bibitem{disk} 
  S.~Pedicelli {\it et al.},
  ``On the metallicity gradient of the Galactic disk,''
  Astron.\ Astrophys.\  {\bf 504}, 81 (2009)
  doi:10.1051/0004-6361/200912504
  [arXiv:0906.3140 [astro-ph.SR]].
 
  \bibitem{53} 
  G.W. Lugmair, A. Shukolyukov, 
``Early solar system timescales according to 53Mn-53Cr systematics", 
Geochimica et Cosmochimica Acta,  {\bf 62}, 2863 (1998)
http://dx.doi.org/10.1016/S0016-7037(98)00189-6.

  N.~Ouellette, S.~J.~Desch and J.~J.~Hester,
  ``Interaction of Supernova Ejecta with Nearby Protoplanetary Disks,''
  Astrophys.\ J.\  {\bf 662}, 1268 (2007)
  doi:10.1086/518102
  [arXiv:0704.1652 [astro-ph]].
 
 \bibitem{al} 
  F.~C.~Adams and G.~Laughlin,
  ``Constraints on the birth aggregate of the solar system,''
  Icarus {\bf 150}, 151 (2001)
  doi:10.1006/icar.2000.6567
  [astro-ph/0011326].

\bibitem{fv}
D. A. Fischer and J. Valenti,
``The Planet-Metallicity Correlation",
Astrophys. J. {\bf 622}, 1102 (2005)
doi:10.1086/428383.

\bibitem{rub} 
  S.~N.~Raymond, A.~M.~Mandell and S.~Sigurdsson,
  ``Exotic Earths: Forming Habitable Worlds with Giant Planet Migration,''
  Science {\bf 313}, 1413 (2006)
  doi:10.1126/science.1130461
  [astro-ph/0609253].

\bibitem{ms}
W. F, McDonough, and S. S. Sun,
 ``Composition of the Earth",
 Chemical Geology {\bf 120}, 223 (1995)
 doi: 10.1016/0009-2541(94)00140-4.

  \bibitem{ctt2} 
  J.~J.~Cowan, F.~K.~Thielemann and J.~W.~Truran,
  ``The R-process and nucleochronology,''
  Phys.\ Rept.\  {\bf 208}, 267 (1991)
  doi:10.1016/0370-1573(91)90070-3.

\bibitem{raby} 
  S.~Raby,
  ``Grand Unified Theories,''
  hep-ph/0608183.

\bibitem{jwst}
J. P. Gardner {\it et. al.},
``The James Webb Space Telescope",
Space Sci. Rev, {\bf 123}, 485 (2006)
doi:10.1007/s11214-006-8315-7.

\bibitem{cheops}
C. Broeg {\it et. al.},
``CHEOPS: A transit photometry mission for ESA's small mission programme",
European Phys. J. of Web Conf. {\bf 47}, (2013)
doi:10.1051/epjconf/20134703005
[arXiv:1305.2270[astro-ph]].

\bibitem{tess}
G. R. Ricker {\it et. al.},
``Transiting Exoplanet Survey Satellite (TESS)",
Proc. SPIE,  {\bf 9143}, (2014)
doi:10.1117/12.2063489
[arXiv:1406.0151[astro-ph]].

\end{thebibliography}
\end{document}